\documentclass[useAMS,usenatbib,usegraphicx]{mn2e}

\usepackage{epsfig,color,float,threeparttable}

\newcommand{\Ha}{\mbox{H}$\alpha$}

\newcommand{\FeI}{\mbox{Fe\,{\sc i}}}
\newcommand{\FeII}{\mbox{Fe\,{\sc ii}}}
\newcommand{\CaII}{\mbox{Ca\,{\sc ii}}}
\newcommand{\LiI}{\mbox{Li\,{\sc i}}}
\newcommand{\ms}{m~s$^{-1}$}
\newcommand{\kms}{km~s$^{-1}$}
\newcommand{\Teff}{$T_{\rm{eff}}$}
\newcommand{\logg}{\mbox{log $g$}}
\newcommand{\met}{$[\rm{Fe/H}]$}
\newcommand{\SR}{$R_{\odot}$}
\newcommand{\SM}{$M_{\odot}$}
\newcommand{\JR}{$R_{J}$}
\newcommand{\JM}{$M_{J}$}
\newcommand{\Gmas}{4.01$\pm$0.35}
\newcommand{\Grad}{$1.49^{+0.16}_{-0.18}$}

\title[Inflated HJ in close orbit around a late F-star]
      {The first planet detected in the WTS: an inflated hot-Jupiter in a 
       3.35 day orbit around a late F-star\thanks{Based on observations collected at  
              the 3.8-m United Kingdom Infrared Telescope (Hawaii, USA),
              the Hobby-Eberly Telescope (Texas, USA),
              the 2.5-m Isaac Newton Telescope (La Palma, Spain),
              the William Herschel Telescope (La Palma, Spain),               
              the German-Spanish Astronomical Center (Calar Alto, Spain), 
              the Kitt Peak National Observatory (Arizona, USA) and
              the Hertfordshire's Bayfordbury Observatory.
              }
      }

\author[M. Cappetta et al.]{M. Cappetta$^{1}$\thanks{E-mail: cappetta@mpe.mpg.de}, 
R.P. Saglia$^{1,2}$, J.L. Birkby$^{3,6}$, J. Koppenhoefer$^{1,2}$, D.J. Pinfield$^{4}$,\and
S.T. Hodgkin$^{3}$, P. Cruz$^{5}$, G. Kov\'acs$^{3}$, B. Sip\H{o}cz$^{4}$, D. Barrado$^{5,16}$, B. Nefs$^{6}$,\and
Y.V. Pavlenko$^{7}$, L. Fossati$^{8}$, C. del Burgo$^{9,10,11}$, E.L. Mart\'in$^{12}$, I. Snellen$^{6}$, \and
J. Barnes$^{4}$, A. M. Bayo$^{18}$, D. A. Campbell$^{4}$, S. Catalan$^{4}$, M.C. G\'alvez-Ortiz$^{12}$, \and
N. Goulding$^{4}$, C. Haswell$^{8}$, O. Ivanyuk$^{7}$, H. Jones$^{4}$, M. Kuznetsov$^{7}$, N. Lodieu$^{13}$, \and
F. Marocco$^{4}$, D. Mislis$^{3}$, F. Murgas$^{13,14}$, R. Napiwotzki$^{4}$,
E. Palle$^{13,14}$, D. Pollacco$^{15}$, \and
L. Sarro Baro$^{17}$, E. Solano$^{5,19}$, P. Steele$^{1}$, H. Stoev$^{5}$, R. Tata$^{13,14}$, J. Zendejas$^{1,2}$ \\
$^1$Max-Planck-Institut f\"ur extraterrestrische Physik, Giessenbachstrasse, 
D-85741 Garching, Germany\\
$^2$Universitatssternwarte Scheinerstrasse 1, D-81679 Munchen, Germany \\
$^3$Institute of Astronomy, University of Cambridge, Madingley Road, Cambridge, 
CB3 0HA, UK \\
$^4$Center for Astrophysics Research, University of Hertfordshire, College Lane, 
Hatfield, Hertfordshire AL10 9AB, UK\\
$^5$Departamento de Astrof\'isica, Centro de Astrobiolog\'ia (CSIC/INTA), 
PO Box 78, E-28691 Villanueva de la Ca\~nada, Spain \\
$^6$Leiden Observatory, Lieiden University, Postbus 9513, 2300 RA, Leiden, 
The Netherlands\\
$^7$Main Astronomical Observatory of Ukrainian Academy of Sciences, Golosiiv Woods, 
Kyiv-127, 03680, Ukraine\\
$^8$Department of Physical Sciences, The Open University, Walton Hall, Milton Keynes, 
MK7 6AA, UK \\
$^{9}$UNINOVA-CA3, Campus da Caparica, Quinta da Torre, Monte de Caparica 2825-149, 
Caparica, Portugal\\
$^{10}$School of Cosmic Physics, Dublin Institute for Advanced Studies, 
Dublin 2, Ireland\\
$^{11}$Instituto Nacional de Astrof\'isica, \'Optica y Electr\'onica (INAOE), Aptdo. 
Postal 51 y 216, 72000 Puebla, Pue., Mexico\\
$^{12}$Centro de Astrobiolog\'ia (CSIC-INTA). Crta, Ajalvir km 4. E-28850, 
Torrej\'on de Ardoz, Madrid, Spain\\
$^{13}$Instituto de Astrof\'isica de Canarias, Calle V\'ia L\'actea s/n, 
E-38200 La Laguna, Tenerife, Spain\\
$^{14}$Departamento de Astrof\'isica, Universidad de La Laguna (ULL), 
E-38205 La Laguna, Tenerife, Spain\\
$^{15}$Astrophysics Research Centre, School of Mathematics \& Physics, 
Queen’s University, University Road, Belfast BT7 1NN\\
$^{16}$Calar Alto Observatory, Centro Astronómico Hispano Alemán, C/ Jesús 
Durbán Remón, E-04004 Almería, Spain\\
$^{17}$Departamento de Inteligencia Artificial, UNED, Juan del Rosal, 
16, 28040 Madrid, Spain\\
$^{18}$European Southern Observatory, Alonso de C\'ordova 3107, Vitacura, Santiago, Chile
$^{19}$Spanish Virtual Observatory\\
}

\begin{document}

\date{Accepted 2012 August 14,  Received 2012 August 10; in original form 2012 July 17}

\pagerange{\pageref{firstpage}--\pageref{lastpage}} \pubyear{2012}

\maketitle

\label{firstpage}

\begin{abstract} 
We report the discovery of WTS-1b, the first extrasolar planet found by the
WFCAM Transit Survey, which began observations at the 3.8-m United Kingdom
Infrared Telescope (UKIRT) in August 2007. Light curves comprising almost 1200
epochs with a photometric precision of better than 1 per cent to $J\sim16$ were
constructed for $\sim60\,000$ stars and searched for periodic transit signals.
For one of the most promising transiting candidates, high-resolution spectra
taken at the Hobby-Eberly Telescope (HET) allowed us to estimate the
spectroscopic parameters of the host star, a late-F main sequence dwarf
(V=16.13) with possibly slightly subsolar metallicity, and to measure its radial
velocity variations. The combined analysis of the light curves and spectroscopic
data resulted in an orbital period of the substellar companion of 3.35 days, a
planetary mass of \Gmas\,\JM, and a planetary radius of \Grad\ \JR. WTS-1b has
one of the largest radius anomalies among the known hot Jupiters in the mass
range 3-5\,\JM. The high irradiation from the host star ranks the planet in the
pM class.
\end{abstract}


\section{Introduction}\label{intro}

The existence of highly-irradiated, gas-giants planets orbiting within $<0.1$ AU
of their host stars, and the unexpected large radii of many of them, is an
unresolved problem in the theory of planet formation and evolution
\citep{Baraffe10}. Their prominence amongst the $777$ confirmed
exoplanets\footnote{http://www.exoplanet.eu at the time of the publication of
this work} is unsurprising; their large radii, large masses, and short orbital
periods make them readily accessible to ground-based transit and radial velocity
surveys, on account of the comparatively large flux variations and reflex
motions that they cause. Exoplanet searches often focus on the detection of
small Earth-like exoplanets, but understanding the formation mechanism and
evolution of the giant planets, particularly in the overlap mass regime with
brown dwarfs, is a key question in astrophysics.

This paper reports on WTS-1b, the first planet discovery from the WFCAM Transit
Survey (WTS; \citealt{Kovacs12,Birkby11}). The WTS is the only large-scale
ground-based transit survey that operates at near infrared (NIR) wavelengths.
The advantage of photometric monitoring at NIR wavelengths is an increased
sensitivity to photons from M-dwarfs ($M_{\star} < 0.6$\,\SM). These small stars
undergo similar flux variations and reflex motions in the presence of an
Earth-like companion as solar-type stars do with hot Jupiter companions. While a
primary goal of the WTS is the detection of terrestrial planets around cool
stars, WTS can also provide observational constraints on the mechanism for giant
planet formation, by accurately measuring the hot Jupiter fraction for M-dwarfs.
The light curve quality of the WTS is sufficient to reveal transiting
super-Earths around mid-M dwarfs \citep{Kovacs12}, but scaling-up means that any
hot Jupiter companions to the $\sim$80\,000 FGK stars in the survey are also
detectable. 

Theoretical models of isolated giant planets predict an almost constant radius
for pure H$+$He objects in the mass range $0.5-10$\,\JM\ as a result of the
equilibrium between the electron degeneracy in the core and the pressure support
in the external gas layers
\citep{ZapolskySalpeter69,Guillot05,Seager07,Baraffe10}. Hence the larger radii
of many hot Jupiters (hereafter HJ) must arise from other factors. Due to the
proximity of these planets to their host star, the irradiation of the surface of
the planet is thought to play a major role in the so-called radius anomaly, by
altering the thermal equilibrium and delaying the Kelvin-Helmholtz contraction
of the planet from birth \citep[e.g.][]{ShowmanGuillot02}. This is supported by
a correlation between the mean planetary density and the incident stellar flux
\citep{Laughlin11,DemorySeager11,Enoch12}. However, it has been shown that this
cannot be the only explanation for the radius anomaly \citep{Burrows07}, and
other sources must contribute to the large amount of energy required to keep gas
giants radii above $\sim$1.2\,\JR\ \citep{Baraffe10}.

There are a number of physical mechanisms thought to be responsible for radius
inflation, including (but not limited to): tidal heating due to the
circularisation of close-in orbits
\citep{Bodenheimer01,Bodenheimer03,Jackson08}, reduced heat loss due to enhanced
opacities in the outer layers of the planetary atmosphere \citep{Burrows07},
double-diffusive convection leading to slower heat transportation
\citep{Chabrier07,LeconteChabrier12}, increased heating via Ohmic dissipation in
which ionised atoms interact with the planetary magnetic field as they move
along strong atmospheric winds \citep{BatyginStevenson10}, and a slower cooling
rate due to the mechanical greenhouse effect in which turbulent mixing drives a
downwards flux of heat simulating a more intense incident irradiation
\citep{YoudinMitchell10}. A radically different explanation has been proposed by
\citet{Martin11} who point out a correlation between radius anomaly and tidal
decay timescale and suggest that inflated HJs are actually young because they
have recently formed as a result of binary mergers. Studying the radius anomaly
in higher mass HJs ($>3$\,\JR) is useful as they are perhaps more resilient to
atmospheric loss due to their larger Roche lobe radius.
   
This paper is organised as follows: in Section~\ref{wts_survey} we briefly
describe the strategy of the WFCAM Transit Survey, while the photometric and
spectroscopic observations are presented in Section~\ref{photo_data} along with
their related data reduction. Section~\ref{analysis} describes how we
characterized the host star (Section~\ref{star}) and determined the properties
of the planet (Section~\ref{planet}), using a combination of low- and
high-resolution spectra and photometric follow-up observations.  A discussion on
the nature and the peculiarity of this new planet, our conclusions and further
considerations are given in Section~\ref{disc}. Throughout this paper we refer
to the planet as WTS-1b, and to the parent star and the whole star-planet system
as WTS-1.

\subsection{The WFCAM Transit Survey}\label{wts_survey}
The WTS is an on-going photometric monitoring campaign using the Wide
Field Camera (WFCAM) on the United Kingdom Infrared Telescope (UKIRT)
at Mauna Kea, Hawaii, and has been in operation since
August 2007. UKIRT is a 3.8-m telescope, designed solely for
NIR observations and operated in queue-scheduled mode. The
survey was awarded 200 nights of observing time, of which $\sim$50$\%$ 
has been observed to-date, and runs primarily as a back-up program when 
observing conditions are not optimal for the main UKIDSS programs 
(seeing $>1$ arcsec).
The survey targets four 1.6 deg$^{2}$ fields, distributed seasonally in 
right ascension at 03, 07, 17 and 19 hours to allow year-round visibility. 
WTS-1b is located in the 19 hour field (hereafter 19hr). The fields are 
close to the Galactic plane but have $b>5$ degrees to avoid over-crowding 
and high reddening. 
The exact locations were chosen to maximise the number of M-dwarfs, while 
keeping giant contamination at a minimum and maintaining an $E(B-V)<0.1$.
Due to the back-up nature of the program, observations of a given field 
are randomly distributed throughout a given night, but on average occur 
in a one hour block at its beginning or end.
Seasonal visibility also leaves long gaps when no observations are
possible \citep{Kovacs12}. Since the WTS was primarily designed to find
planets transiting M-dwarf stars, the observations are obtained in
the J-band ($\sim$1.25\,$\mu$m). This wavelength is near to the peak of
the spectral energy distribution (SED) of a typical M-dwarf. Hotter stars,
with an emission peaking at shorter wavelength, are consequently
fainter in this band. Interestingly, photometric monitoring at NIR 
wavelengths may have a further advantage as it is potentially less 
susceptible to the effect of star-spot induced variability \citep{Goulding12}.  
The discovery reported in this paper demonstrates that despite the 
survey optimisation for M-dwarfs, its uneven epoch distribution 
and the increased difficulty in obtaining high-precision light curves from
ground-based infrared detectors, it is still able to detect HJs around FGK stars.

\section{Observations}\label{obs}  
\subsection{Photometric data}\label{photo_data} 
\subsubsection{UKIRT/WFCAM $J$-band photometry}\label{wfcam} 

\begin{figure*}
\centering
\includegraphics[width=1.0\textwidth]{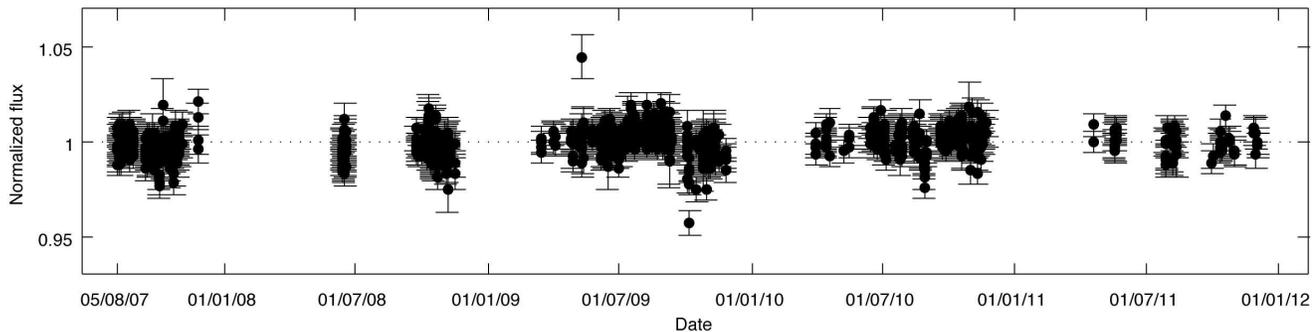}
\caption{The unfolded $J$-band light curve for WTS-1.}
\label{plot_RAWJ}
\end{figure*}

\begin{table}
\caption{WFCAM $J$-band light curve. The full epoch list, which
         contains 1\,182 entries, will be available electronically.}
  \centering
    \begin{tabular}{lcc}
    \hline
    HJD            & Normalized &  Error         \\
    -2\,400\,000   & flux       &      \\
    \hline      
    54317.8138166 & 1.0034 & 0.0057 \\
    54317.8258565 & 0.9971 & 0.0056 \\
    ... & ... & ... \\
    55896.7105702 & 0.9979 & 0.0064 \\
    \hline \\
    \end{tabular}    
   \label{tab_Jband}
\end{table}

The instrument used for this campaign is WFCAM \citep{Casali07}, which has a
mosaic of four Rockwell Hawaii-II PACE infrared imaging 2048 $\times$ 2048
pixels detectors, covering $13.65\arcmin\times13.65\arcmin$ ($0.4\arcsec$/pixel)
each. The detectors are placed in the four corners of a square (this pattern is
called a paw-print) with a separation of $12.83\arcmin$ between the chips,
corresponding to $94\%$ of a chip width. Each of the four fields consists of
eight pointings of the WFCAM paw-print, each one comprising a 9-point jitter
pattern of 10 second exposures each, and tiled to give uniform coverage across
the field. It takes 15 minutes to observe an entire WTS field ($9 \times 10s
\times 8$ + overheads). In this way, the NIR light-curves have an average
cadence of 4 data points per hour.

A full description of our 2-D image processing and light curve generation is
given by \citet{Kovacs12} and closely follows that of \citet{Irwin07}. Briefly,
a modified version of the CASU INT wide-field survey pipeline
\citep{IrwinLewis01}
\footnote{http://casu.ast.cam.ac.uk/surveys-projects/wfcam/technical/} is used
to remove the dark current and reset anomaly from the raw images, apply a
flat-field correction using twilight flats, and to decurtain and sky subtract
the final images. Astrometric and photometric calibration is performed using
2MASS stars in the field-of-view \citep{Hodgkin09}. A master catalogue of source
positions is generated from a stacked image of the 20 best frames and
co-located, variable aperture photometry is used to extract the light curves. We
attempt to remove systematic trends that may arise from flat-fielding
inaccuracies or varying differential atmospheric extinction across the image by
fitting a 2-D quadratic polynomial to the flux residuals in each light curve as
a function of the source's spatial position on the chip. As a final correction,
for each source we remove residual seeing-correlated effects by fitting a
second-order polynomial to the correlation between its flux and the stellar
image FWHM per frame (see \citealt{Irwin07} for a discussion of the
effectiveness of these techniques).

The final $J$-band light curves have a photometric precision of 1$\%$ down to
$J=16$\,mag ($\sim$7$\%$ for $J=18$\,mag while the uncertainty per data point is
just 3 mmag for the brightest stars (saturation occurs between
$J\sim12-13$\,mag). The unfolded $J$-band light curve for WTS-1 is shown in
{\mbox{Figure}}~\ref{plot_RAWJ} (all the measurements are reported in
{\mbox{Table}}~\ref{tab_Jband}) and its final out-of-transit RMS is 0.0064
(equivalent to 6.92\,mmag). 

\subsubsection{Transit detection algorithm}\label{detect} 

\begin{table}  
  \caption{The host star WTS-1.}
  \centering
  \begin{threeparttable}[b]
  {
  \renewcommand{\arraystretch}{1.3}
  \begin{tabular}{rcl}
    \hline    
    Parameter & & Value \\ 
    \hline        
    RA\tnote{a}          && 19h 35m 56.4s                                     \\
    Dec\tnote{a}         && +36d 17m 46.8s                                    \\
    l\tnote{a}           && +70.0162\,deg                                    \\
    b\tnote{a}           && +7.5573\,deg                                     \\
    $\mu_{\alpha} cos\delta$\tnote{b}  && -6.1$\pm$1.9\,mas~yr$^{-1}$                \\
    $\mu_{\delta}$\tnote{b}         && -2.8$\pm$2.4\,mas~yr$^{-1}$                   \\
    \hline \\
  \end{tabular}
  }
  \begin{tablenotes}    
    \item [a] Epoch J2000;
    \item [b] Proper motion from SDSS.
  \end{tablenotes}
\end{threeparttable}
\label{tab_WTS1_star}
\end{table} 

We identified WTS-1b in the $J$-band light curves by
using the Box-Least-Squares transit search algorithm {\sc occfit},
as described in \citet{AigrainIrwin04}, which takes a maximum
likelihood approach to fitting generalized periodic step functions.
Before inspecting transit candidates by eye, we employed several
criteria to speed up the detection process. The first is a magnitude
cut, in which we removed all sources fainter than $J=17$\,mag. We also
required that the source have an image morphology consistent with
stellar sources \citep{Irwin07}. Despite our attempts to remove
systematic trends in the light curves, we invariably suffered from
residual correlated red noise, so we modified the detection
significance statistic, $S$, from {\sc occfit} according to the
prescription (equation 4) of \citet{Pont06}, to obtain $S_{red}$.
Our transit candidates must then have $S_{red}\ge5$ to survive, although
we note that this is more permissive with respect to the limit
recommended by Pont and collaborators ($S_{red}\ge7$). We further discarded any
detections with a period in the range $0.99<$P$<1.005$ days, in
order to avoid the common $\sim$1 day alias of the ground-based
photometric surveys.

Next, as fully described in \citet{Birkby12a}, the WFCAM $Z Y J H K$ single 
epoch photometry was combined with five more optical photometric data points 
($u g r i z$ bands) available for the 19hr field from the Sloan Digitized 
Sky Survey archive \citep[SDSS $7^{th}$ release,][]{Nash96} to create an 
initial SED. The SEDs were
fitted with the NextGen models \citep{Baraffe98} in order to
estimate effective temperatures and hence a stellar radius for each
source. We could then impose a final threshold on the detected
transit depths, by rejecting those that corresponded to a planetary
radius greater than 2\,\JR. It is worth noting that {\sc occfit}
tends to under-estimate transit depths because it does not consider
limb-darkening effects and the trapezoidal shape of the transits.
Moreover, the NextGen models systematically under-estimate the
temperature of solar-like stars \citep{Baraffe98}, so the initial
radius estimates are also under-predicted. Hence, the genuine HJ
transit events are unlikely to be removed in this final threshold cut.
As a result, our transits detection procedure was conservative. Many
of the $\sim$3\,500 phase-folded light curves which satisfied our
criteria were false-positives arising from nights of bad data or
single bad frames. Others were binary systems, folded
on half the true orbital period. A more detailed analysis of the
candidate selection procedure and of advanced selection steps in the
survey can be found in \citet{Sipocz12}. 

The WTS-1 $J$-band light curve passed all our selection criteria and the 
object (see {\mbox{Table}}~\ref{tab_WTS1_star}) progressed to the following 
phases of candidate's confirmation. 
The descriptions of the photometric and spectroscopic data in the 
next sections are organized in order to match the following chronological 
sequence of analysis.
First, optical $i'$-band photometric follow-up of WTS-1 was conduced 
in order to prove the real presence of the transit and check the 
transit depth consistency between the two bands.
Then, ISIS/WHT intermediate resolution spectra enabled us to place strong constrains 
on the velocity amplitude of host star (at the \kms level) and therefore to 
rule out the false-positive eclipsing binaries scenarios. Finally we moved to the
high-resolution spectroscopic follow-up.
The confirmation of the planetary nature of WTS-1b came with the RV 
measurements obtained with the HET spectra.

\subsubsection{Broad band photometry}\label{bbp} 

The WFCAM and SDSS photometric data for WTS-1 are reported in 
{\mbox{Table}}~\ref{tab_BBP} with other single epoch broad band photometric observations.
Johnson $B$ $V$ $R$ bands were observed for WTS-1 on the night of 6th
April 2012 at the University of Hertfordshire's Bayfordbury Observatory. 
We used a Meade LX200GPS 16-inch f/10 telescope fitted with an SBIG STL-6303E CCD camera,
and integration times of 300 seconds per band. Images were bias, dark, and
flat-field corrected, and extracted aperture photometry was calibrated
using three bright reference stars within the image. Photometric
uncertainties combine contributions from the SNR of the source
(typically $\sim$20) with the scatter in the zero point from the calibration
stars. 
The Two Micron All Sky Survey \citep[2MASS,][]{Skrutskie06} and the Wide-field 
Infrared Survey Explorer \citep[WISE,][]{Wright10} provide further 
NIR data points ($J$ $H$ $Ks$ bands and $W1$ $W2$ bands respectively).

\begin{table}
\caption{Broad band photometric data of WTS-1 measured within the WFCAM (Vega), 
             SDSS (AB), 2MASS (Vega) and WISE (Vega) surveys. Johnson magnitudes
             in the visible are provided too (Vega).
             Effective wavelength $\lambda_{eff}$ (mean wavelength weighted by 
             the transmission function of the filter), equivalent width (EW) 
             and magnitude are given for each single pass-band.
             The bands are sorted by increasing $\lambda_{eff}$.}
  \centering
    \begin{tabular}{lrrcc}
    \hline 
    Band & $\lambda_{eff}$[\AA] & EW[\AA] & Magnitude \\ 
    \hline 
    SDSS-u    & 3546  & 558  & 18.007 ($\pm$0.014) \\
    Johnson-B & 4378  & 970  & 17.0   ($\pm$0.1) \\
    SDSS-g    & 4670  & 1158 & 16.785 ($\pm$0.004) \\
    Johnson-V & 5466  & 890  & 16.5   ($\pm$0.1) \\
    SDSS-r    & 6156  & 1111 & 16.434 ($\pm$0.004) \\
    Johnson-R & 6696  & 2070 & 16.1   ($\pm$0.1) \\
    SDSS-i    & 7471  & 1045 & 16.249 ($\pm$0.004) \\
    UKIDSS-Z  & 8817  & 879  & 15.742 ($\pm$0.005)\\
    SDSS-z    & 8918  & 1124 & 16.189 ($\pm$0.008) \\
    UKIDSS-Y  & 10305 & 1007 & 15.642 ($\pm$0.007)\\
    2MASS-J   & 12350 & 1624 & 15.375 ($\pm$0.052) \\
    UKIDSS-J  & 12483 & 1474 & 15.387 ($\pm$0.005)\\
    UKIDSS-H  & 16313 & 2779 & 15.103 ($\pm$0.006)\\
    2MASS-H   & 16620 & 2509 & 15.187 ($\pm$0.081) \\
    2MASS-Ks  & 21590 & 2619 & 15.271 ($\pm$0.199) \\
    UKIDSS-K  & 22010 & 3267 & 15.048 ($\pm$0.009)\\
    WISE-W1   & 34002 & 6626 & 15.041 ($\pm$0.044) \\
    WISE-W2   & 46520 & 10422& 15.886 ($\pm$0.157)\\
    \hline \\
    \end{tabular}    
   \label{tab_BBP}
\end{table}

\subsubsection{INT $i'$-band data}\label{inti} 

In addition to the WFCAM $J$-band light curve, we observed one half
transit of the WTS-1 system in the $i'$-band using the Wide Field 
Camera on the 2.5-m Isaac Newton Telescope \citep{Mcmahon01} on July 23, 2010.
A total of 82 images, sampling the ingress of the transit, were taken
with an integration time of 60 seconds. The data were reduced with
the CASU INT/WFC data reduction pipeline as described in detail by
\citet{IrwinLewis01} and \citet{Irwin07}.  The pipeline performs a
standard CCD reduction, including bias correction, trimming of the
overscan and non-illuminated regions, a non-linearity correction,
flat-fielding and defringing, followed by astrometric and
photometric calibration. A master catalogue for the $i'$-band filter
was then generated by stacking 20 frames taken under the best
conditions (seeing, sky brightness and transparency) and running
source detection software on the stacked image. The extracted source
positions were used to perform variable aperture photometry on all
of the images, resulting in a time-series of differential
photometry. 

The final out-of-transit RMS in the WTS-1 $i'$-band light
curve is 0.0026 (equivalent to 2.87\,mmag) and is used to refine the transit model fitting 
procedure in Section~\ref{transit}. The $i'$-band light curve for WTS-1 is given 
in {\mbox{Table}}~\ref{tab_iband}.

\begin{table}
\caption{INT $i'$-band light curve of WTS-1. The full epoch list, which
         contains 82 entries, will be available electronically.}
  \centering
    \begin{tabular}{lcc}
    \hline
    HJD            & Normalized &  Error         \\
    -2\,400\,000   & flux       &      \\
    \hline      
    55401.3703254 & 1.0244 & 0.0109 \\
    55401.3809041 & 1.0205 & 0.0034 \\
    ... & ... & ... \\
    55401.4894461 & 0.9936 & 0.0022 \\
    \hline \\
    \end{tabular}    
   \label{tab_iband}
\end{table}

\subsection{Spectroscopic data}\label{spectra_data} 

\subsubsection{ISIS/WHT}\label{isis} 

We carried out intermediate-resolution spectroscopy of the star 
WTS-1 over two nights between July $29-30$, 2010, as part of a wider
follow-up campaign of the WTS planet candidates, using the William
Herschel Telescope (WHT) at Roque de Los Muchachos, La Palma. We used
the single-slit Intermediate dispersion Spectrograph and Imaging
System (ISIS). The red arm with the R1200R grating centred on 8500\,\AA\ 
was employed. We did not use the dichroic during the ISIS observations
because it can induce systematics and up to $10\%$ efficiency losses
in the red arm, which we wanted to avoid given the relative faintness
of our targets. The four spectra observed have a wavelength coverage of
8100--8900\,\AA. The wavelength range was chosen to be optimal for
the majority of the targets for our spectroscopic observation which
were low-mass stars. The slit width was chosen to match the
approximate seeing at the time of observation giving an average
spectral resolution $R\sim9000$. 
An additional low-resolution spectrum was taken on July 16th, 2010, using 
the ISIS spectrograph with the R158R grating centred on 6500\,\AA.
This spectrum has a resolution ($R\sim1000$), a SNR of $\sim40$ and a 
wider wavelength coverage (5000--9000\,\AA).
The spectra were processed using the {\sc iraf.ccdproc}\footnote[3]{
{\sc iraf} is distributed by National Astronomy Observatories, which is operated
by the Association of Universities of Research in Astronomy, Inc., under
contract to the National Science, USA} package for 
instrumental signature removal. We optimally extracted the spectra and 
performed wavelength calibration using the semi-automatic 
{\sc kpno.doslit} package. The dispersion function employed in the 
wavelength calibration was performed using CuNe arc lamp spectra 
taken after each set of exposures.

\subsubsection{CAFOS/2.2-m Calar Alto}\label{cafos} 

Two spectra of WTS-1 were obtained with CAFOS at the 2.2-m telescope at the Calar Alto 
observatory (as a Director’s Discretionary Time - DDT) in June, 2011. CAFOS is a 
2k${\times}$2k CCD SITe{\#}1d camera at the RC focus, and it was equipped with the 
grism R-100 that gives a dispersion of ${\sim}$ 2.0 \AA/pix and a wavelength coverage 
from 5850 to 9500 \AA, approximately. Their resolving power is of around $R\sim1900$ 
at 7500 \AA, with a $SNR\sim25$. The data reduction was performed  
following a standard procedure for CCD processing and spectra extraction with {\sc iraf}. 
The spectra were finally averaged in order to increase the SNR.

\subsubsection{KPNO}\label{kpno} 

A low resolution spectrum of WTS-1 was observed in September 2011 with the 
Ritchey-Chretien Focus Spectrograph at the 4-m telescope at Kitt Peak (Arizona, USA). 
The grism BL-181, which gives a dispersion of ${\sim}$ 2.8 \AA/pix, was used. 
Calibration, sky subtraction, wavelength and flux calibration were performed following 
a standard procedure for long slit observations using dedicated {\sc iraf} tasks.
The ThAr arc lamp and the standard star spectra, employed for the wavelength 
and the flux calibration respectively, were taken directly after the science exposure.
The measured spectrum covers the wavelength range 6000-9000\,\AA\ with SNR$\sim$40
at 7500\,\AA\ and has a resolution of $R\sim1000$.
This is the only flux calibrated spectrum we have available.

\subsubsection{HET}\label{het} 

In the late 2010 and in the second half of the 2011 the star WTS-1 was 
observed during 11 nights with the High Resolution Spectrograph (HRS) housed in 
the insulated chamber in the basement of the 9.2-m segmented mirror Hobby-Eberly 
Telescope \citep[HET, see][]{Ramsey98}. 
The HRS is a single channel adaptation of the ESO UVES spectrometer linked 
to the corrected prime focus of the HET through its fiber-fed instrument  
as described by \citet{Tull98}. It uses an R-4 echelle mosaic, which we used with 
a resolution of R=60\,000, and a cross-dispersion grating to separate spectral orders, 
while the detector is a mosaic of two thinned and anti-reflection coated 2K $\times$ 4K CCDs. 

One science fiber was used to get the spectrum of the target star and two 
sky fibers were used in order to subtract the sky contamination.
A couple of ThAr calibration exposures were taken immediately before 
and after the science exposure. This strategy allowed us to keep under control 
any undesired systematic effect during the observation. Each science observation 
(except one, see Section~\ref{rv}) 
was split in two exposures, of about half an hour each, in order 
to limit the effect of the cosmic rays hits, which can affect the data 
reduction and analysis steps. Due to the faintness of the star, the Iodine 
gas cell was not used for our observations. 

The data reduction was performed with the {\sc iraf.echelle} package 
\citep{WillmarthBarnes94}. After the standard calibration of the raw science 
frame, bias-subtraction and flat-fielding, the stellar spectra were extracted 
order by order and the related sky spectrum was subtracted.
The extracted ThAr spectra were used to compute the dispersion functions, 
which are characterized by an RMS of the order of 0.003 \AA. 
Consistency between the two solutions (computed from the ThAr 
taken before and after the science exposure) was checked for all the 
nights in order to detect possible drifts or any other technical hitch that 
could take place during each run. Successively, the spectra were wavelength calibrated 
using a linear interpolation of the two dispersion functions.

In the subsequent data analysis, custom Matlab programs were used.
After the continuum estimation and the following normalization, the spectra 
were filtered to remove the residual cosmic rays peaks left after the previous
filtering performed on the raw science frame with the {\sc iraf} task 
{\sc cosmicrays}. Comparing the spectra of all the nights for each single order, 
the pixels with higher flux, due to a cosmic hit on the detector, were detected 
and masked. The telluric absorption lines present in the redder orders were then 
removed in order to avoid contaminations using a high SNR observed spectrum of a 
white dwarf as template for the telluric lines.
Finally, the spectra related to the two split science 
exposures were averaged to obtain the final set of spectra used in the following 
analysis. 
A total of 40 orders (18 from the red CCD and 22 from the blue one) cover the 
wavelength range 4400-6300 \AA. The spectra have a SNR $\sim8$. 

\section{Analysis and results}\label{analysis}  
  
\subsection{Stellar parameters}\label{star} 
    
\subsubsection{Spectral energy distribution}\label{sed} 

\begin{figure}
\centering
\includegraphics[width=0.485\textwidth]{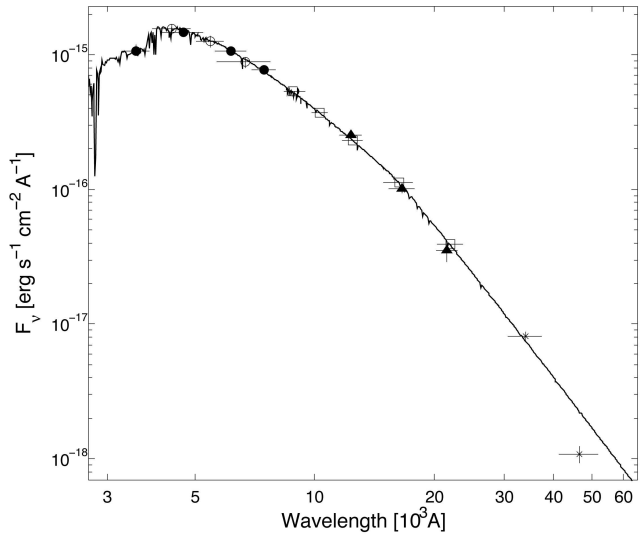}
\caption{Broad band photometric data of WTS-1: SDSS (filled dots), Johnson (empty dots), 
         WFCAM (squares), 
         2MASS (triangles) and WISE (stars). The best fitting template 
         (black line) is the ATLAS9 Kurucz \Teff=6500\,K, \logg=4.5, 
         \met=-0.5 model with $A_{V}$=0.44. Vertical errorbars correspond to the 
         flux uncertainties while those along the X-axis represent each 
         band's EW (see {\mbox{Table}}~\ref{tab_BBP}).}
\label{plot_VOSA}
\end{figure}

A first characterization of the parent star can be performed comparing the shape 
of the Spectral Energy Distribution (SED), constructed from broad band photometric 
observations, with a grid of synthetic theoretical spectra.
The data relative to the photometric bands, collected in {\mbox{Table}}~\ref{tab_BBP}, 
were analysed with the application VOSA 
\citep[Virtual Observatory SED Analyser,][]{Bayo08,Bayo12}.
VOSA offers a valuable set of tools for the SED analysis, allowing the 
estimation of the stellar parameters. It can be accessed through 
its web-page interface and accepts as input file an ASCII table with 
the following data: source identifier, coordinates of the source, distance 
to the source in parsecs, visual extinction ($A_{V}$), filter label, observed 
flux or magnitude and the related uncertainty. 
For our purpose, we tried to estimate only the main intrinsic parameters 
of the star: effective temperature, surface gravity and metallicity. 
The extinction $A_{V}$ was assumed as a further free parameter.

The synthetic photometry is calculated by convolving the response curve of 
the used filter set with the theoretical synthetic spectra. Then a statistical 
test is performed, via $\chi^{2}_{\nu}$ minimization, to estimate which set of 
synthetic photometry best reproduces the observed data. We decided to employ 
the Kurucz ATLAS9 templates set \citep{Castelli97} to fit our photometric data.
These templates reproduce the SEDs in the high temperature regime better than  
the NextGen models \citep{Baraffe98}, more suitable at lower temperatures ($<4500\,K$).
 
In order to speed up the fitting procedure, we restricted the range 
of \Teff\ and \logg\ to 3500-8000\,K and 3.0-5.0, respectively. 
These constrains in the parameter space did not affect the final results as
it was checked a posteriori that the same results were obtained considering the full 
available range for both parameters.
The resulting best fitting synthetic template, plotted in {\mbox{Figure}}~\ref{plot_VOSA} 
with the photometric data, corresponds to the \Teff=6500\,K, \logg=4.5 and 
\met=-0.5 Kurucz model and A$_{V}$=0.44. 
Uncertainties on the parameters were estimated both using $\chi^{2}_{\nu}$ 
statistical analysis and a bayesian (flat prior) approach. The related errors 
result to be of the order of 250\,K, 0.2, 0.5 and 0.07 for \Teff, \logg, \met\ 
and A$_{V}$, respectively.

\begin{figure*}
\centering
\includegraphics[width=0.99\textwidth,angle=0]{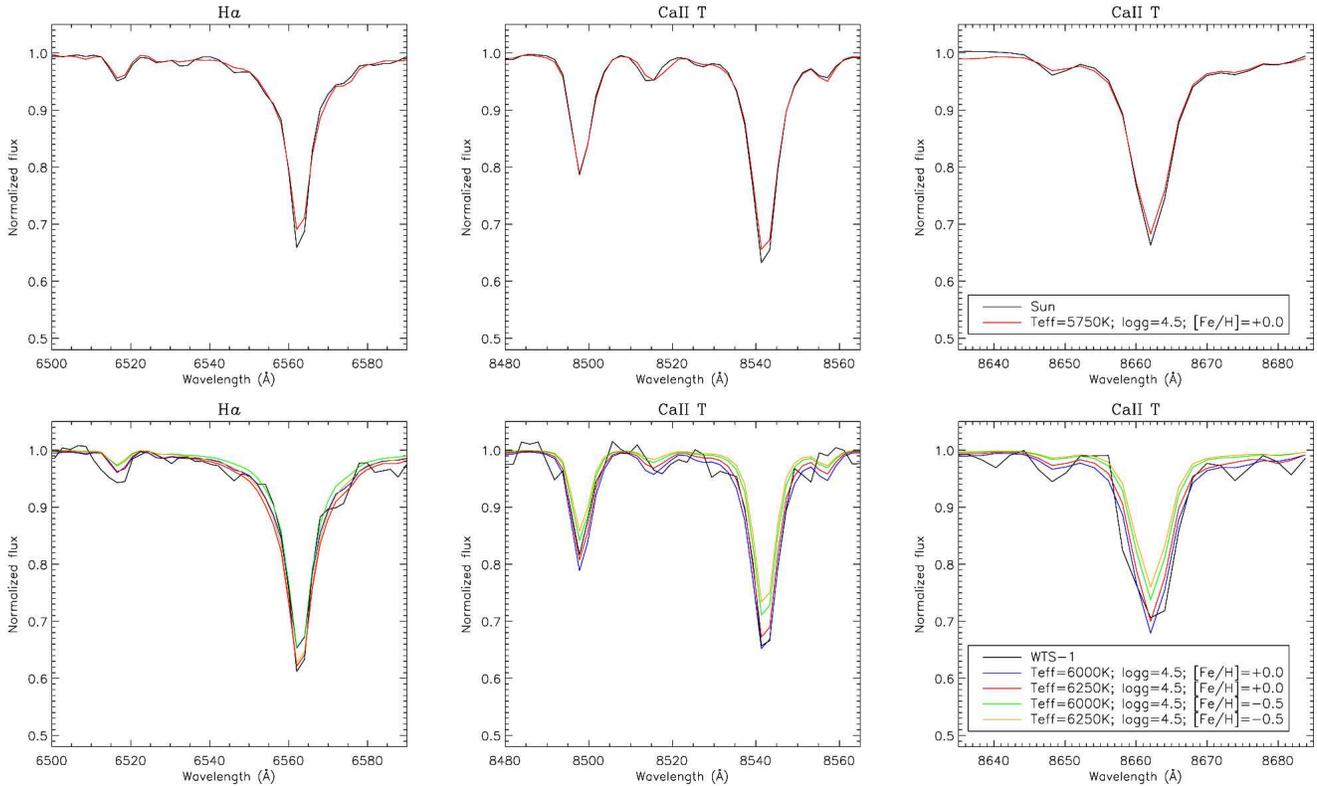}
\caption{$Upper$ $row$: comparison of a degraded spectrum of the Sun (black) 
         with a synthetic spectrum (red) computed with \Teff=5750\,K, \logg=4.5 
         and \met=0.0. $Lower$ $row$: comparison of the WTS-1 spectrum (black) 
         with different synthetic spectra (colours).
         The comparison between the observed WTS-1 spectrum and the synthetic models 
         took into account the differences of the core of the lines shown in 
         the upper row plots.  
         From this analysis, the best fitting model is the one with \Teff=6250\,K, 
         \logg=4.5 and \met=0.0.}
\label{plot_sunfit}
\end{figure*}

As it can be seen in {\mbox{Figure}}~\ref{plot_VOSA}, the WISE $W2$ data point is not 
consistent
with the best fitting model. Firstly, it is worth noting that the observed value of
$W2=15.886$ ($\pm$0.157) is below the 5 sigma point source sensitivity expected in
the $W2$-band ($>15.5$). The number of single source detections used for the $W2$-band 
measurement is also considerably less than that of the $W1$-band (4 and 19 respectively) 
increasing the uncertainty in the measurement. Finally, the poor angular resolution of WISE
in the $W2$-band ($6.4\arcsec$) could add further imprecisions to the final measured flux,
especially in a field as crowded as the WTS 19hrs field. For these reasons, the WISE $W2$ data
point was not considered in the fitting procedure.

Once the magnitude values were corrected for the interstellar absorption according 
to the best fitting value (A$_{V}$=0.44$\pm$0.07), colour -- temperature relations were 
used to further check the effective temperature and spectral type of the host star. 
From the SDSS $g$ and $r$ magnitudes, we obtained a value of 
(B-V)$_{0}$=0.43$\pm$0.04 \citep{Jester05} which imply \Teff=6300$\pm$600\,K assuming 
\logg=4.4 and \met=-0.5 \citep{SekiFuku00}. Following the appendix B of \citet{CCameron07}, 
the 2MASS ($J-H$) index of 0.23$\pm$0.09 leads to a value of \Teff=6200$\pm$400\,K while
{\mbox{Table}} 3 of \citet{Covey07} suggests the host star to be a late-F considering 
different colour indices at once. These results are all compatible within the uncertainties.

\subsubsection{Spectroscopic analysis}\label{spectra} 

The spectroscopic spectral type determination was done firstly by comparing the spectrum 
observed with CAFOS with a set of spectra of template stars. Stars of different spectral 
types, uniformly spanning the F5 to G2 range, were observed with the same instrumental 
setting. 
Since the observed spectrum has a relatively low SNR, we focused the 
analysis on the strongest features present which are the \Ha\ (6562.8\,\AA) and 
the \CaII\ triplet (8498.02\,\AA, 8542.09\,\AA, 8662.14\,\AA). The best match was obtained, 
via minimization of the RMS of the difference between the WTS-1 and template star spectra,
with the spectrum of an F6V star with solar metallicity.

Afterwards, we tried to estimate the stellar parameters comparing the observed 
spectrum with a simulated spectrum with known parameters. 
For that aim, we used a library of high resolution synthetic stellar 
spectra by \citet{Coelho05}, created by the PFANT code \citep{Barbuy82,Cayrel91,
Barbuy03} that computes a synthetic spectrum assuming local thermodynamic equilibrium 
(LTE). The synthetic spectra were achieved using the model atmospheres presented by 
\citet{CastelliKurucz03}.
Since the core of these lines are strongly affected by cumulative effects of 
the chromosphere, non-LTE (local thermodynamical equilibrium) and inhomogeneity of 
velocity fields, we firstly compared
a spectrum of the Sun observed with HIRES spectrograph at the Keck telescope 
\citep{Vogt94}. The spectrum of the Sun was degraded to lower resolution and resampled 
to match our CAFOS spectrum specifications to see 
how such effects appear at this resolution and how different the solar spectrum is from 
a synthetic spectrum from the library by Coelho and collaborators. 
Looking at the upper plots in {\mbox{Figure}}~\ref{plot_sunfit}, we concluded that
the cores of the lines in the simulated spectrum are systematically higher than 
those in the observed spectrum of the Sun. Nevertheless, the \CaII\ triplet line 
at 8498.02\,\AA\ seems to be less affected by the above-mentioned problems.
Considering these differences between central depth of the observed and simulated 
spectra in the best fitting procedure, we estimated that the synthetic model with 
\Teff=6250\,K, \logg=4.5 and \met=0.0 best reproduces the observed spectrum of WTS-1. 
The expected uncertainties are of the order of the step size of the used library 
($\delta$\Teff=250\,K, $\delta$\logg=0.5 and $\delta$\met=0.5).

\begin{figure}
\centering
\includegraphics[width=0.47\textwidth]{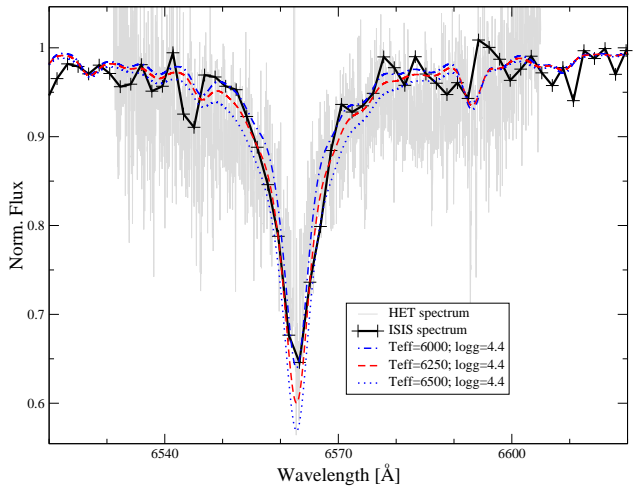}
\caption{Intermediate resolution ISIS spectrum (black line) and high resolution 
HET spectrum (grey line) of the \Ha\ line of WTS-1 in comparison with synthetic spectra 
calculated with different effective temperatures of 6000\,K
(blue dash-dotted line), 6250\,K (dashed line; finally adopted \Teff), and
6500\,K (blue dotted line). The synthetic spectra were convolved in order to match 
the resolution of the ISIS spectrum.}
\label{plot_het_isis}
\end{figure}

The high resolution HET spectra were employed to attempt a more detailed 
spectroscopic analysis of the host star. The spectra observed each single 
night were stacked together obtaining a final spectrum with a SNR of about 12,
calculated over a 1\,\AA\ region at 5000\,\AA. 
To compute model atmospheres, LLmodels stellar model atmosphere code 
\citep{Shulyak04} was used. For all the calculations, LTE and plane-parallel geometry 
were assumed. 
We used the VALD database \citep{Piskunov95,Kupka99,Ryabchikova99} as a source 
of atomic line parameters for opacity calculations with the LLmodels code. 
Finally, convection was implemented according to the \citet{CanutoMazzitelli91} 
model of convection.

The parameter determination and abundance analysis were performed 
iteratively, self-consistently recalculating a new model atmosphere any time 
one of the parameters, including the abundances, changed. As a starting
point, we adopted the parameters derived from the CAFOS spectrum. We 
performed the atmospheric parameter determination by imposing the iron 
excitation and ionization equilibria making use of equivalent widths measured
for all available unblended and weakly blended lines. We converted the
equivalent width of each line into an abundance value with a modified version 
\citep{Tsymbal96} of the WIDTH9 code \citep{Kurucz93}. Unfortunately, 
the faintness of the observed star, coupled with the calibration process 
(including the sky subtraction), led to a distortion of the stronger
lines, weakening their cores. For this reason, our analysis took into account only the
measurable weak lines, making therefore impossible a determination of the
microturbulence velocity ($v_{mic}$), which we fixed at a value of 0.85\,\kms\
\citep{ValentiFisher05}. With the fixed $v_{mic}$, we imposed the Fe 
excitation and ionization equilibria, which led us to an effective temperature 
of 6000$\pm$400\,K and a surface gravity of 4.3$\pm$0.4. Imposing the 
ionization equilibrium we took into account the expected non-LTE effects for 
\FeI\ \citep[$\sim$0.05 dex,][]{Mashonkina11}, while for \FeII\ non-LTE 
effects are negligible. 

In this temperature regime, the ionization equilibrium is sensitive to both 
\Teff\ and \logg\ variations, therefore it is important to simultaneously and
independently further constrain temperature and/or gravity. For this reason 
we looked at the \Ha\ line to further constrain \Teff, as in this temperature 
regime \Ha\ is sensitive primarily to temperature variations \citep{Fuhrmann93}. 
We did this by fitting synthetic spectra, calculated with SYNTH3 
\citep{Kochukhov07}, to the \Ha\ line profile observed in the HET high resolution 
and in the ISIS/WHT low resolution spectra.
As the \Ha\ line of the HET spectrum was also affected by the before mentioned
reduction problems, only the wings of the line were considered. Although we
could calculate synthetic line profiles of \Ha\ on the basis of atmospheric
models with any \Teff, because of the low SNR of our spectra, we performed the
line profile fitting on the basis of models with a temperature step of 100\,K
\citep[small variations in gravity and metallicity are negligible]{Fuhrmann93}.
From the fit of the \Ha\ line we obtained a best fitting \Teff\ of
6100$\pm$400\,K. Further details on method, codes and techniques can be found in
\citet{Fossati09}, \citet{Ryabchikova09}, \citet{Fossati11} and references
therein. {\mbox{Figure}}~\ref{plot_het_isis} shows the \Ha\ line profile
observed with ISIS and HET in comparison with synthetic spectra calculated
assuming a reduced set of stellar parameters. On the basis of the previous
analysis, we finally adopted Teff=6250$\pm$250\,K, log g=4.4$\pm$0.1. With this
set of parameters and the equivalent widths measured with the HET spectrum, a
final metallicity of -0.5$\pm$0.5\,dex dex was derived, where the uncertainty
takes into account the internal scatter and the uncertainty on the atmospheric
parameters. By fitting synthetic spectra, calculated with the final atmospheric
parameters and abundances, to the observation, we derived a projected rotational
velocity v $sin(i)$=7$\pm$2\,\kms.

\begin{figure}
\centering
\includegraphics[width=0.47\textwidth]{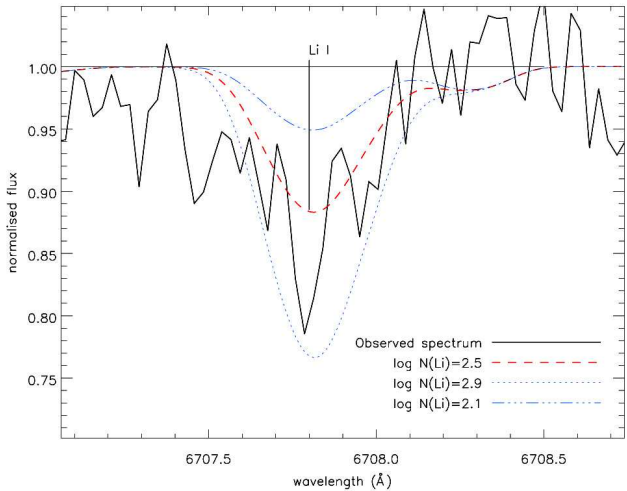}
\caption{High-resolution HET stacked spectrum (black full line) of the WTS-1 \LiI\ line
at $\sim$6707\,\AA\ in comparison with three synthetic spectra calculated with 
the final adopted stellar parameters and lithium abundances ($\log$ N(Li)) 
of 2.5 (red dashed line), 2.9 (blue dotted line), and 2.1 (blue dash-dotted 
line). The vertical line indicates the position of the centre of the 
multiplet.}
\label{lithium}
\end{figure}

The HET spectrum allowed us to measured the atmospheric lithium abundance from
the \LiI\ line at $\sim$6707\,\AA. Lithium abundance is important as it can constrain 
the age of the star (see Section~\ref{wts1}). As this line presents a strong hyperfine
structure and is slightly blended by a nearby \FeI\ line, we measured the \LiI\
abundance by means of spectral synthesis, instead of equivalent widths. By
adopting the meteoritic/terrestrial isotopic ratio Li6/Li7=0.08 by 
\citet{RosmanTaylor98}, we derived $\log$ N(Li)=2.5$\pm$0.4 (corresponding to an 
equivalent width of 41.12$\pm$24.40\,m\AA), where the uncertainty takes into 
account the uncertainty on the atmospheric parameters, \Teff\ in particular 
(see {\mbox{Figure}}~\ref{lithium}). 

The low resolution spectrum obtained at the KPNO observatory, 
being the only flux calibrated spectrum, was compared to the Kurucz 
ATLAS9 templates set (the same employed in the SED analysis in Section~\ref{sed}).
The limited wavelength range covered by the observed spectrum (3000\,\AA), 
allowed to achieve usable results fitting only one parameter of the template spectra. 
We therefore decided to leave \Teff\ as a free parameter, fixing the other quantities
to \logg=4.5, \met=-0.5 and A$_{V}$=0.44.
We concluded that the temperature range 6000-6500\,K brackets the \Teff\ 
with 95$\%$ confidence level.

\subsubsection{Properties of the host star WTS-1}\label{wts1} 

The atmospheric parameters of the star WTS-1 were computed combining the results
coming from the analysis described in the previous sections. We finally obtained
an effective temperature of 6250$\pm$200\,K, a surface gravity of 4.4$\pm$0.1
and a metallicity range $-0.5\div0.0$\,dex. As described in
Section~\ref{spectra}, the faintness of the star and the reduction process led
to a distortion of the stronger lines in the HET high resolution spectrum,
reducing the number of reliable lines employed in the measure of the
metallicity. For these reasons, the relative uncertainty on the metallicity is
larger with respect to those of the other parameters. Further observations would
be needed to pin down the exact value of the star metallicity.

In order to determine the parameters of the stellar companion, mass and radius
of the host star must be known. The fit of the transit in the light curves
provides important constrains on the mean stellar density (see
Section~\ref{transit}). Joining this quantity to the effective temperature, we
could place WTS-1 in the modified $\rho_{s}^{-1/3}$ vs. \Teff\ H-R diagram and
compare its position with evolutionary tracks and isochrones models
\citep{Girardi00} in order to estimate stellar mass and age. In this way, we
estimated the stellar mass to be 1.2$\pm$0.1\,\SM\ and the age of the system to
range between 200\,Myr and 4.5\,Gyr. 

Further constrains on the stellar age can be fixed considering the measured
\LiI\ line abundance. Depletion of lithium in stars hotter than the Sun is
thought to be due to a not yet clearly identified slow mixing process during the
main-sequence evolution, because those stars do not experience pre-main sequence
depletion \citep{Martin97}. Comparison of the lithium abundance of WTS-1 ($\log$
N(Li)=2.5$\pm$0.4) with those of open clusters rise the lower limit on the age
to 600\,Myr, because younger clusters do not show lithium depletion in their
late-F members \citep{SestitoRandich05}. On the other hand, it is not feasible
to derive an age constrain from the WTS-1 rotation rate (v
$sin(i)$=7$\pm$2\,\kms) because stars with spectral types earlier than F8 show
no age-rotation relation \citep{Wolff86} and thus they are left out from the
formulation of the spin down rate of low-mass stars \citep{Stepien88}. 

The true space motion knowledge of WTS-1 allows to determine to which component
of our Galaxy it belongs. In order to compute the U, V and W components of the
motion, the following set of quantities was required: distance, systemic
velocity (from the RV fit, see Section~\ref{rv}), proper motion \citep[from SDSS
$7^{th}$ release,][]{Munn04,Munn08} and coordinates of the star. The distance to
the observed system was estimated according to the extinction, fitted in the SED
analysis (A$_{V}$=$0.44\pm0.07$), and a model of dust distribution in the galaxy
\citep[][axis-symmetric model]{AmoresLepine05}. The UVW values and their errors
are calculated using the method in \citet{JohnsonSoderblom87}, with respect to
the Sun (heliocentric) and in a left-handed coordinate system, so that they are
positive away from the Galactic centre, Galactic rotation and the North Galactic
Pole respectively. All the quantities here discussed are listed in
{\mbox{Table}}~\ref{tab_WTS1}. Considering the uncertainties on the derived
quantities, maily affected by the error on the distance, the host star is
consistent with both the definitions of Galactic young-old disk and young disk
populations \citep[metallicity between -0.5 and 0\,dex, solar-metallicity
respectively,][]{Leggett92}. But we could assess that WTS-1 is not a halo
member.

Combining $\rho_s$ and M$_{s}$, a value of $1.15^{+0.10}_{-0.12}$\,\SR\ was
computed for the stellar radius. The same result was obtained considering the
stellar mass and the surface gravity measured from the spectroscopic analysis.
Scaling by the distance the apparent $V$ magnitude calculated from the SDSS $g$
and $r$ magnitudes \citep{Jester05}, we computed an absolute $V$ magnitude of
$3.55^{+0.27}_{-0.38}$. This value and all the other derived stellar parameters
are consistent with each other and with the typical quantities expected for an
F6-8 main-sequence star \citep{Torres09}. They are collected in
{\mbox{Table}}~\ref{tab_WTS1}.

\begin{table}  
  \caption{Properties of the WTS-1 host star.}
  \centering
  \begin{threeparttable}[b]
  {
  \renewcommand{\arraystretch}{1.3}
  \begin{tabular}{rcl}
    \hline    
    Parameter & & Value \\ 
    \hline    
    \Teff\               && 6250$\pm$200\,K                                  \\
    \logg\               && 4.4$\pm$0.1                                  \\
    \met\                && [-0.5, 0]\,dex                               \\    
    M$_s$                && 1.2$\pm$0.1 \SM\  \\
    R$_s$                && $1.15^{+0.10}_{-0.12}$ \SR\  \\    
    m$_V$                && 16.13$\pm$0.04                                  \\
    M$_V$                && $3.55^{+0.27}_{-0.38}$                                  \\
    v $sin(i)$\tnote{a}  && 7$\pm$2\,\kms                                    \\
    $\rho_s$             && $0.79^{+0.31}_{-0.18}$ $\rho_{\rm{sun}}$              \\
    Age                  && [0.6, 4.5]\,Gyr                                 \\
    Distance             && $3.2^{+0.9}_{-0.4}$\,kpc                                  \\
    True space motion U\tnote{c}    && 13$\pm$28\,\kms                            \\
    True space motion V\tnote{c}    && 20$\pm$38\,\kms                             \\
    True space motion W\tnote{c}    && -12$\pm$26\,\kms                            \\    
    \hline \\
  \end{tabular}
  }
  \begin{tablenotes}    
    \item [a] We assumed $v_{mic}$=0.85\,\kms;
    \item [b] Epoch J2000;
    \item [c] Left-handed coordinates system (see text).
  \end{tablenotes}
\end{threeparttable}
\label{tab_WTS1}
\end{table}

\subsection{Planetary parameters}\label{planet} 
    
\subsubsection{Transit fit}\label{transit} 

The light curves in $J$- and $i'$-band were fitted with analytic models
presented by \citet{MandelAgol02}. We used quadratic limb-darkening 
coefficients for a star with effective temperature \Teff=6250\,K, surface 
gravity \logg=4.4 and metallicity \met=-0.5\,dex, calculated as linear interpolations 
in \Teff, \logg\ and \met\ of the values tabulated in \citet{ClaretBloemen11}. 
We use their table derived from {\mbox{ATLAS}}
atmospheric models using the flux conservation method (FCM) which gave
a slightly better fit than the ones derived using the least-squares
method (LSM). The values of the limb-darkening coefficients we used in
our fitting are given in {\mbox{Table}}~\ref{tab_limbcoeff}.
Scaling factors were applied to the error values of the $J$- and
$i'$-band light curves (0.94 and 0.9 respectively) in order to achieve a 
reduced $\chi^2$ of the constant out-of-transit part equal to 1.

\begin{figure}
\begin{center}
\includegraphics[width=0.49\textwidth]{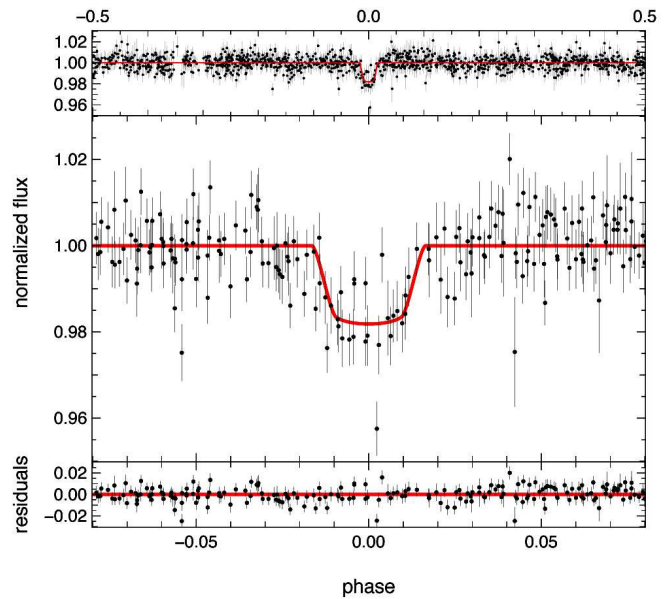}
\caption{WFCAM J-band light curve data of WTS-1. 
         $Upper$ $panel$: whole set of the folded WFCAM J-band data points.
         $Middle$ $panel$: folded photometric data centred in the transit 
         and best model (red line) fitted in combination with the INT i$'$-band data.
         $Lower$ $panel$: residuals of the best fit.}
\label{plot_LCJ}
\end{center}
\end{figure}

\begin{figure}
\centering
\includegraphics[width=0.48\textwidth]{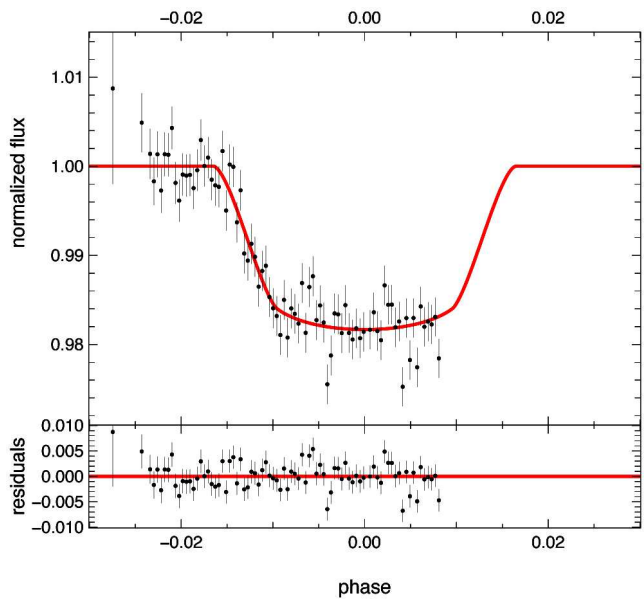}
\caption{INT $i'$-band light curve data of WTS-1. 
         $Upper$ $panel$: photometric data and best model 
         (red line) fitted in combination with the WFCAM $J$-band data.
         $Lower$ $panel$: residuals of the best fit.}
\label{plot_LCi}
\end{figure}

Using a simultaneous fit to both light curves, we fitted the period $P$, 
the time of the central transit t$_0$, the radius ratio $R_{\rm{p}}/R_{\rm{s}}$,
the mean stellar density, $\rho_s = M_{\rm{s}}/R^3_{\rm{s}}$ in solar units and the 
impact parameter $\beta_{\rm{impact}}$ in units of $R_{\rm{s}}$. 
The light curves and the model fit are shown in {\mbox{Figures}}~\ref{plot_LCJ} and 
\ref{plot_LCi}, while the resulting parameters are listed in {\mbox{Table}}~\ref{tab_LCfit}.

\begin{table}
\caption{Quadratic limb-darkening coefficients used for the transit fitting, 
    for a star with effective temperature \Teff=6250\,K, surface gravity \logg=4.4 
    and metallicity \met=-0.5\,dex \citep{ClaretBloemen11}.}
  \centering  
  \begin{tabular}{ccc}
  \hline 
  filter  & $\gamma_1$ & $\gamma_2$  \\ 
  \hline
  J       & 0.14148    & 0.24832     \\ 
  i$'$    & 0.25674    & 0.26298     \\
  \hline \\
  \end{tabular}    
  \label{tab_limbcoeff}
\end{table}

\begin{figure*}
\centering
\includegraphics[width=0.245\textwidth]{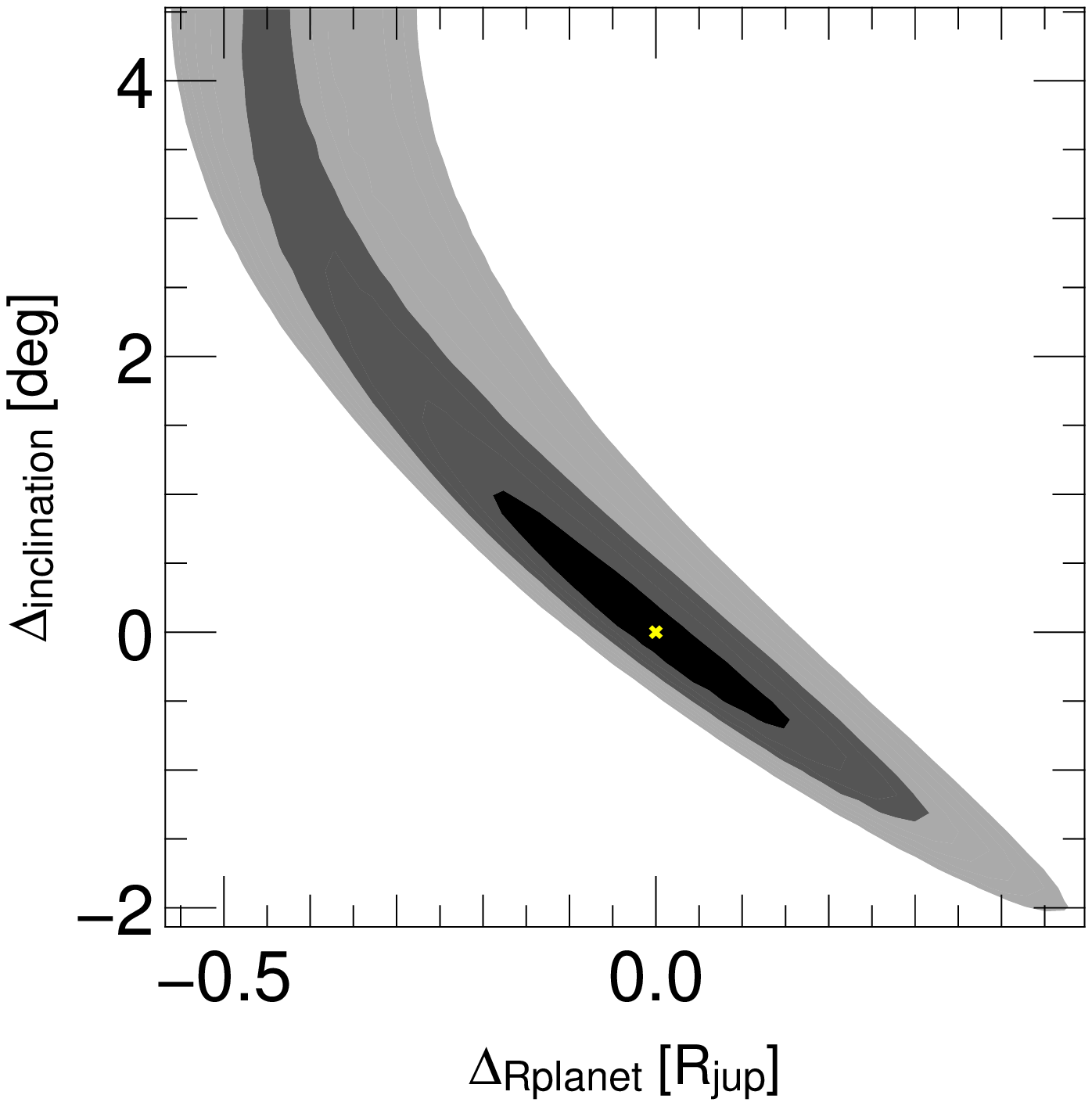}
\includegraphics[width=0.245\textwidth]{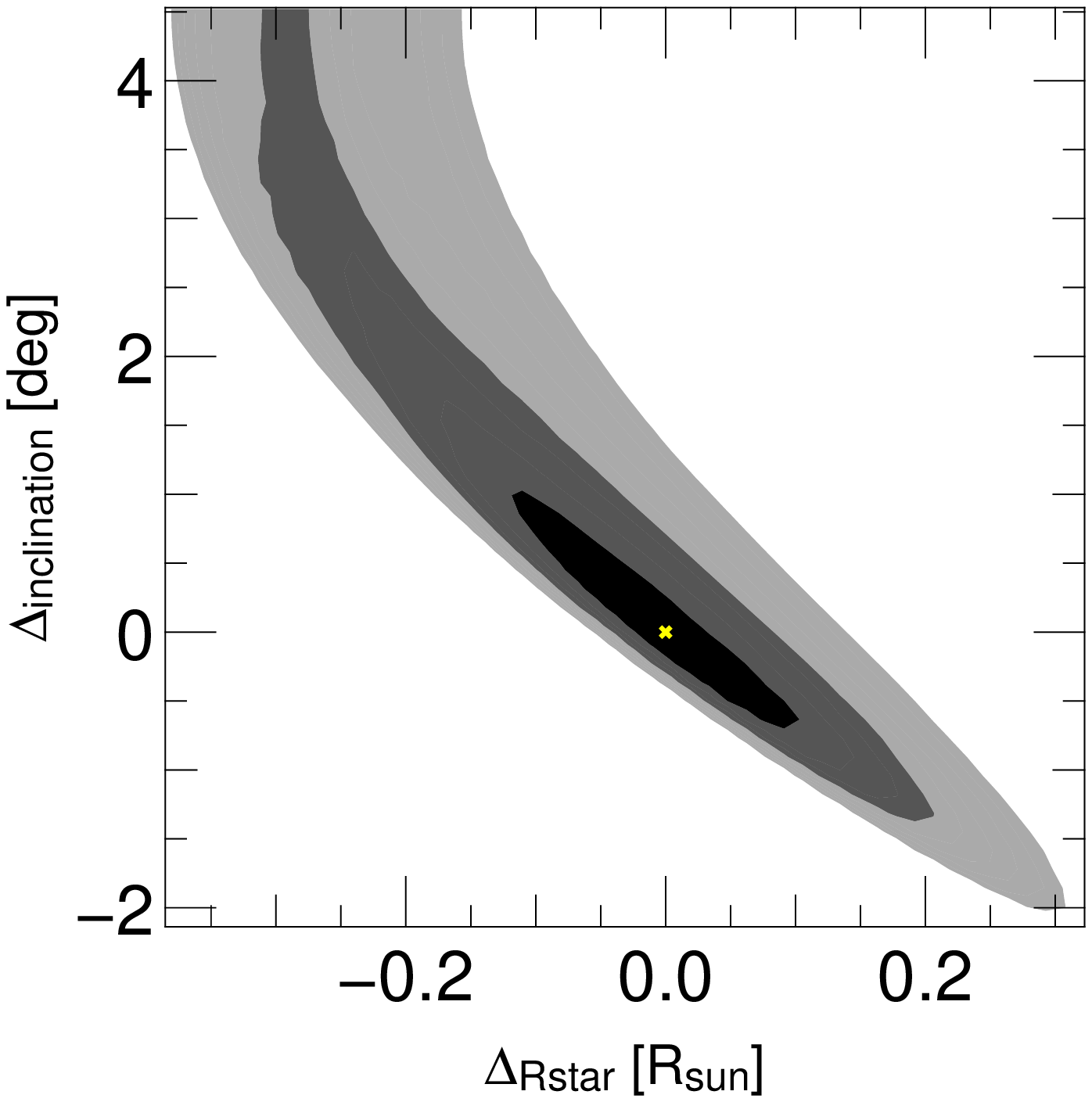}
\includegraphics[width=0.245\textwidth]{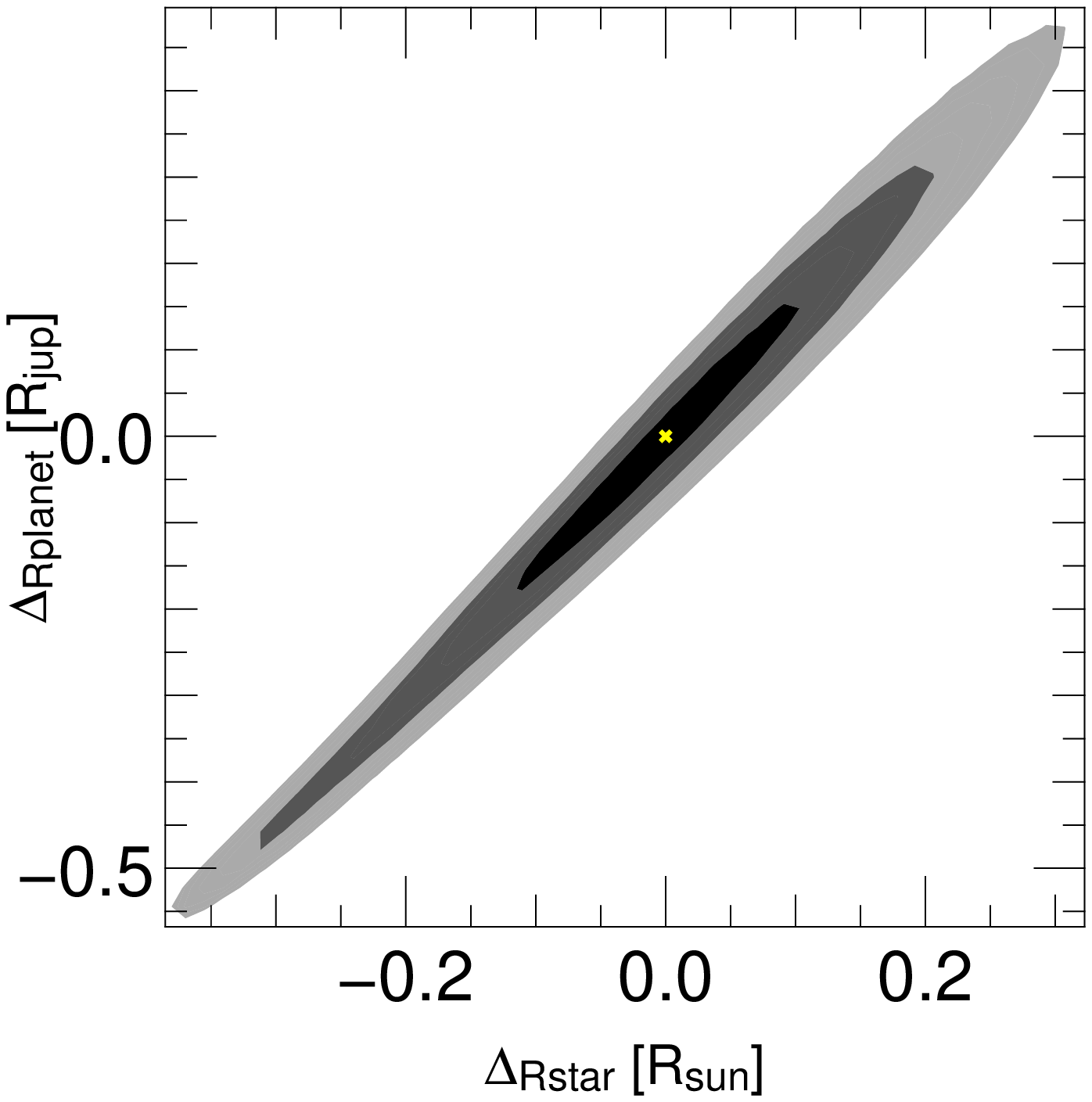}
\includegraphics[width=0.245\textwidth]{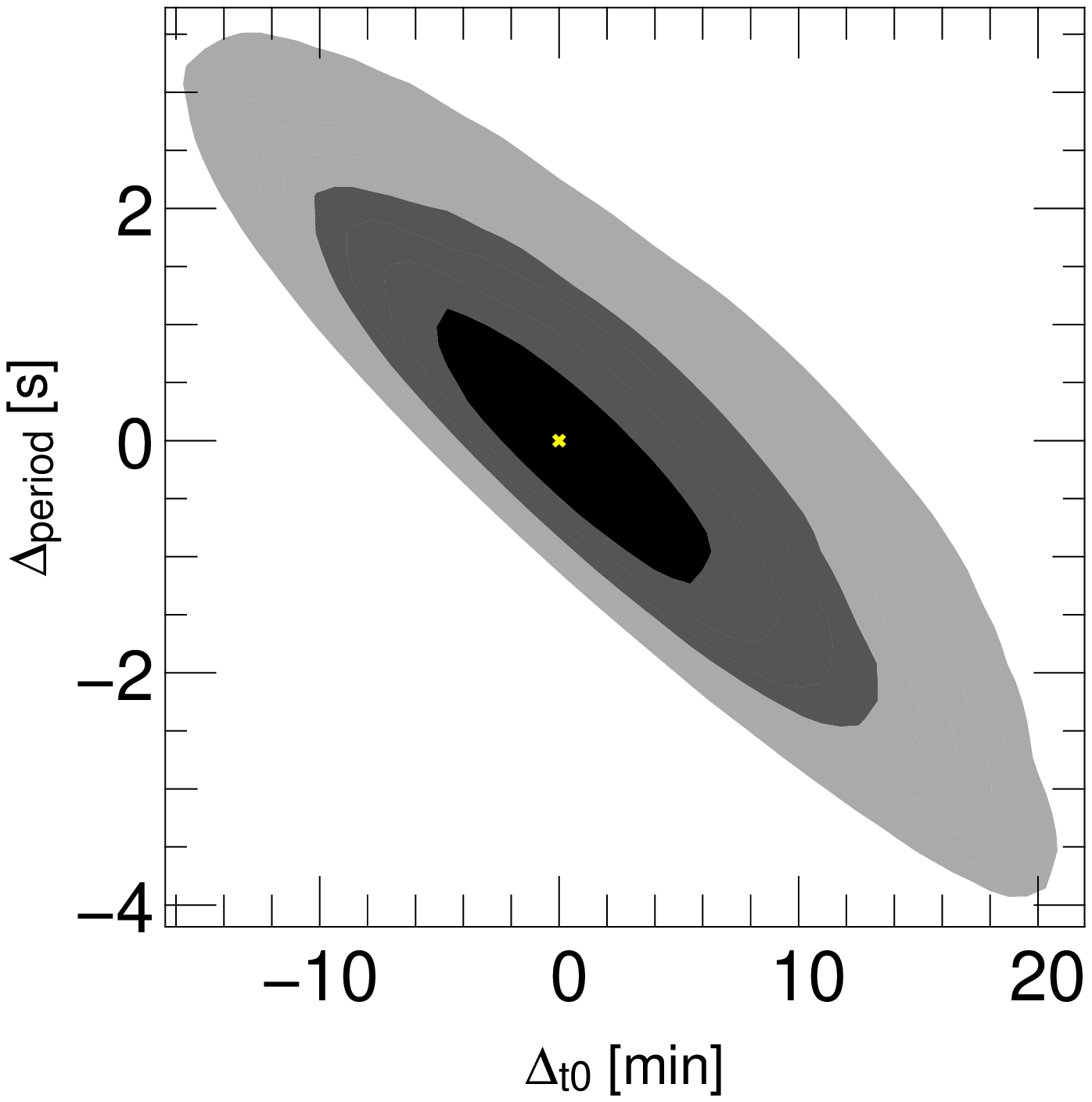}
\caption{Correlation plots of the quantities derived from the simultaneous fit
         of the transit in the $J$ and $i'$-band light curves. 
         The $\chi^{2}$ minima are indicated by crosses while the different tones of grey
         correspond to the 68$\%$, 95$\%$ and 99$\%$ confidence level (darker to 
         lighter respectively).
         The other couples of parameters do not show significant correlations.}
\label{plot_paramcorr}
\end{figure*}

The errors were calculated using a multi-dimensional grid on which we search for
extreme grid points with $\Delta\chi^2$=1 when varying one parameter and
simultaneously minimizing over the others. {\mbox{Figure}}~\ref{plot_paramcorr}
shows the correlations between the parameters of the WTS-1 system derived from
the simultaneous fit of the $J$- and $i'$-band light curves. Considering the
solar metallicity scenario, with different limb-darkening coefficients, the
change in the final fitting parameters is smaller than 1\% of their
uncertainties. Note that the combined fit assumes a fixed radius ratio although
in a hydrogen-rich atmosphere, molecular absorption and scattering processes
could result in different radius ratios in each band \citep[an attempt to detect
such variations has recently been undertaken by][]{Mooij12}. In our case, the
uncertainties are too large to see this effect in the light curves and the
assumption of a fixed radius ratio is a good approximation. The estimated
stellar density of the host star ($0.79^{+0.31}_{-0.18}$ $\rho_{\rm{sun}}$) is
consistent with the expected value based on its spectral type \citep{Seager03}.
The noise in the data did not allow a secondary transit detection.

Subsequently, we searched for further periodic signals in the light curve
after the removal of the data points related to the transit. 
No significant signals were detected in the Lomb-Scargle periodogram up to
a period of 400 days.
Since WTS-1 is a late F-star, there are not many spots on the surface and 
they do not live long enough to produce a stable signal 
over a timescale of several years.

\begin{table}
\caption{Fitted parameters of the WTS-1 system as determined from the simultaneous 
    fit of the $J$- and $i'$-band light curves. Scaling factors to the uncertainties 
    of the $J$ and $i'$ data points (0.94 and 0.9 respectively) were applied in order 
    to achieve a reduced $\chi^2$=1 in the constant out-of-transit part of the light 
    curves.}
  \centering  
  {
  \renewcommand{\arraystretch}{1.5}
  \begin{tabular}{rcl}
    \hline
    Parameter & & Value \\
    \hline   
    $P_{orb}$             &=& $3.352059^{+1.2\times10^{-5}}_{-1.4\times10^{-5}}$ days \\
    t$_0$                &=& 2\,454\,318.7472$^{+0.0043}_{-0.0036}$ HJD \\
    R$_p$/R$_s$          &=& $0.1328^{+0.0032}_{-0.0035}$ \\
    $\rho_s$             &=& $0.79^{+0.31}_{-0.18}$ $\rho_{\rm{sun}}$ \\
    $\beta_{\rm{impact}}$   &=& $0.69^{+0.05}_{-0.09}$       \\
    \hline \\
  \end{tabular}
  }
  \label{tab_LCfit}
\end{table}

\subsubsection{Radial velocity}\label{rv} 

One of the most pernicious transit mimics in the WTS are eclipsing
binaries. On one hand, a transit can be mimicked by an eclipsing
binary that is blended with foreground or background star. 
On the other hand, grazing eclipsing binaries with near-equal
radius stars also have shallow, near-equal depth eclipses that can
phase-fold into transit-like signals at half the binary orbital
period. In order to rule out the eclipsing binaries scenarios, the RV 
variation of the WTS-1 system were first measured using the ISIS/WHT 
intermediate resolution spectra. Eclipsing binaries systems typically 
show RV amplitudes of tens of \kms, while the measured RVs were all 
consistent with a flat trend within the RV uncertainties of $\sim1$\,\kms.

\begin{table}
\caption{Radial velocities and bisector spans measurements for WTS-1 obtained by HET spectra. 
         The phases were computed from the epochs of the observations expressed in Julian date
         and using the $P$ and $t_{0}$ values found with the transit fit.}
  \centering  
  {
  \renewcommand{\arraystretch}{1.2}
  \begin{tabular}{cccc}
    \hline
    HJD            & Phase & RV     & BS       \\
    -2\,400\,000   &       & [\kms] & [\kms]   \\
    \hline   
    55477.676  &  0.74  &  $-1.46\pm0.12$  &  $-0.31\pm0.40$  \\
    55479.666  &  0.33  &  $-1.80\pm0.10$  &  $-0.77\pm0.39$  \\
    55499.608  &  0.28  &  $-2.03\pm0.10$  &  $ 0.76\pm0.51$  \\
    55513.586  &  0.45  &  $-2.06\pm0.17$  &  $ 0.87\pm0.63$  \\
    55522.556  &  0.13  &  $-2.09\pm0.44$  &  $-0.39\pm0.35$  \\
    55523.542  &  0.42  &  $-1.70\pm0.16$  &  $-0.13\pm0.44$  \\
    55742.729  &  0.26  &  $-1.97\pm0.15$  &  $ 0.27\pm0.68$  \\
    55760.673  &  0.79  &  $-1.11\pm0.12$  &  $-0.47\pm0.59$  \\
    55782.837  &  0.22  &  $-2.23\pm0.07$  &  $ 0.41\pm0.39$  \\
    55824.722  &  0.25  &  $-2.31\pm0.06$  &  $ 0.16\pm0.56$  \\
    55849.650  &  0.75  &  $-1.19\pm0.06$  &  $ 0.18\pm0.31$  \\ 
    \hline
  \end{tabular}
  }  
  \label{tab_RVhet}
\end{table}

\begin{figure}
\centering
\includegraphics[width=0.49\textwidth]{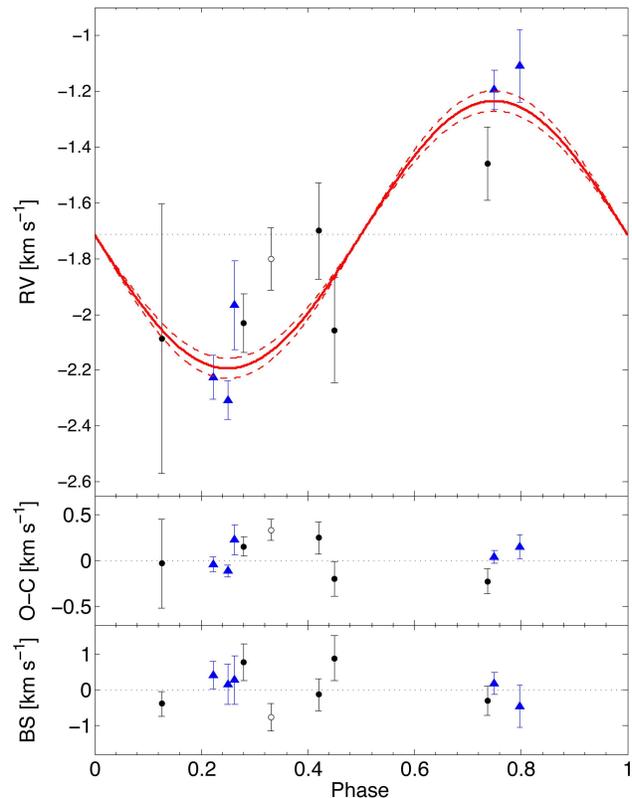}
\caption{$Top$ $panel$: RVs values measured with the high-resolution HET spectra of 
WTS-1 as a function of the orbital phase. 
Black dots and blue triangles refers to observations 
performed in 2010 and 2011 respectively. The data point at $\phi=0.33$, empty dot, 
was excluded from the fitting procedure (see text for details). 
Best-fit circular orbit model ($\chi^{2}_{\nu}$=1.45) 
and 1$\sigma$ uncertainty of the semi-amplitude ($K_{\star}$=479$\pm$34\,\ms) are indicated 
with solid and dashed red lines respectively. The black dotted line refers to the fitted 
radial systemic velocity ($\gamma$=-1\,714$\pm$35\,\ms).
Zero phase corresponds to the mid-transit time.  
$Middle$ $panel$: Phase folded O-C residuals from the best-fit. The residual 
scatter is of the order of $\sim150$\,\ms, consistent with the RVs uncertainties. 
$Bottom$ $panel$: Bisector spans. No significant deviations from zero and no correlation
with the RVs were found.}
\label{plot_RVhet}
\end{figure}

Afterwards, we analysed the high-resolution HET spectra in order to 
accurately investigate the properties of the sub-stellar companion of WTS-1.
The spectra related to each single night were cross-correlated using the {\sc iraf.rv.fxcorr} 
task with the synthetic spectrum of a star with \Teff=6250\,K, \logg=4.4 and \met=-0.5. 
Changes in the effective temperature of the synthetic templates, even of the order of 
several hundreds of Kelvin, cause variations of the measured RV values smaller than the 
statistical uncertainties due to noise in the spectra (suggesting 
the absence of contamination of back/foreground stars of different spectral type). 
Even smaller variations occur changing surface gravity and metallicity of the template.
The 40 single RV values, each one coming from a different order, were used to compute 
the RVs and related uncertainty at each epoch of the observations.
Resampling statistical tools were used in order to better estimate mean value, standard 
deviation and possible bias in the sample of measured RVs. 
Finally, the measured RVs were corrected for the Earth orbital movements and reduced to 
the heliocentric rest-of-frame.
The phase values $\phi$ were computed from the epochs of the observations, 
expressed in Julian date, and using the extremely well determined $P$ and $t_{0}$
values (relative uncertainties are of the order of $10^{-6}$ and $10^{-9}$ 
respectively) obtained from the photometric fit (see {\mbox{Table}}~\ref{tab_LCfit}).

The data, listed in {\mbox{Table}}~\ref{tab_RVhet}, were then fitted with a simple two 
parameters sinusoid of the form:

\begin{equation}
RV=\gamma+K_{\star}sin(2\pi\phi)
\end{equation}

where $K_{\star}$ is the RV semi-amplitude of the host star and $\gamma$ is the 
systemic velocity of the system.
Thanks to the acquisition of two ThAr calibration exposures (before and after the 
science exposure, see Section~\ref{het}), we detected a small drift between the 
ThAr lines occurring during the science exposures on November 22, 2010. 
In the presence of a suspected systematic trend, which could affect the measured RV value, 
we performed the fitting procedure excluding the data point related to that night 
(at $\phi=0.33$). 
The larger RV error of the data point at $\phi=0.13$ is due to the integration 
time (half hour) of the science frame which is shorter than those of all the other data 
points (one hour).
The best fitting model ($\chi^{2}_{\nu}$=1.45) was obtained for $K_{\star}$=479$\pm$34\,\ms
and $\gamma$=-1\,714$\pm$35\,\ms and is plotted in {\mbox{Figure}}~\ref{plot_RVhet} with 
the RV data. 
We imposed the orbit to be circular as the eccentricity was compatible with zero when a 
Keplerian orbit fit was performed (see \citet{Anderson12} for a discussion of 
the rationale for this). In accordance to the RV uncertainties, a relatively 
loose upper limit can be plaved on the eccentricity ($e<0.1$, C.L.$=95\%$).
The fitted RV semi-amplitude implies a planet mass of \Gmas\ \JM\, assuming a host 
star mass of 1.2$\pm$0.1 \SM\ (see Section~\ref{wts1}). The uncertainty on the planet 
mass in mainly driven by the uncertainty on the mass of the host star.
As can be seen in {\mbox{Figure}}~\ref{plot_RVhet}, the RVs related to observations
performed in late 2010 and in the second half of 2011 are consistent, showing no 
significant long term trends in our measurements.   

The HET spectra were employed also to investigate the possibility that the measured 
RVs are not due to true Doppler motion in response to the presence of a planetary 
companion. Similar RV variations can rise in case of distortions in the line profiles 
due to stellar atmosphere oscillations \citep{Queloz01}.
To assert that this is not our case, we used the same cross-correlation profiles
produced previously for the RV calculation to compute the bisector spans 
(BS hereafter) which are reported in {\mbox{Table}}~\ref{tab_RVhet}.
Following \citet{Torres05}, we measured the difference between the bisector 
values at the top and at the bottom of the correlation function for the different 
observation epochs. In case of contaminations, we would have expected to measure 
BS values consistently different from zero and a strong correlation with the measured 
RVs \citep{Queloz01,Mandushev05}.  
As it can be seen in the bottom panel of {\mbox{Figure}}~\ref{plot_RVhet}, the measured BS do 
not show significant deviation from zero within the uncertainties. No correlation 
was detected between the BS and the RV values. In this way, contaminations
that could mimic the effect of the presence of a planet were ruled out.


\section{Discussion and conclusions}\label{disc}  

\begin{table}  
  \caption{Properties of the new extrasolar planet WTS-1b.}
  \centering
  \begin{threeparttable}[b]
  {
  \renewcommand{\arraystretch}{1.3}
  \begin{tabular}{rcl}
    \hline
    Parameter & & Value \\  
    \hline
    M$_p$                && \Gmas\ \JM  \\
    R$_p$                && \Grad\ \JR  \\     
    $P_{rot}$             && $3.352057^{+1.3\times10^{-5}}_{-1.5\times10^{-5}}$ d    \\
    a                    && 0.047$\pm$0.001 AU                               \\
    e                    && $<0.1$ (C.L.$=95 \%$)                            \\ 
    inc                  && $85.5^{+1.0}_{-0.7}$  deg                          \\ 
    $\beta_{\rm{impact}}$   && $0.69^{+0.05}_{-0.09}$                        \\ 
    t$_0$                && 2\,454\,318.7472$^{+0.0043}_{-0.0036}$ HJD           \\
    $\rho_p$             && 1.61$\pm$0.56 $g~cm^{-3}$, 1.21$\pm$0.42 $\rho_{J}$           \\
    $T_{eq}$\tnote{a}     && 1500$\pm$100\,K           \\     
    \hline \\
  \end{tabular}
  }
  \begin{tablenotes}
  \item [a] We assumed a Bold albedo A$_{B}$=0 and re-irradiating fraction F$=1$;    
  \end{tablenotes}
\end{threeparttable}
\label{tab_WTS1b}
\end{table}     

\begin{figure}
\centering
\includegraphics[width=0.49\textwidth]{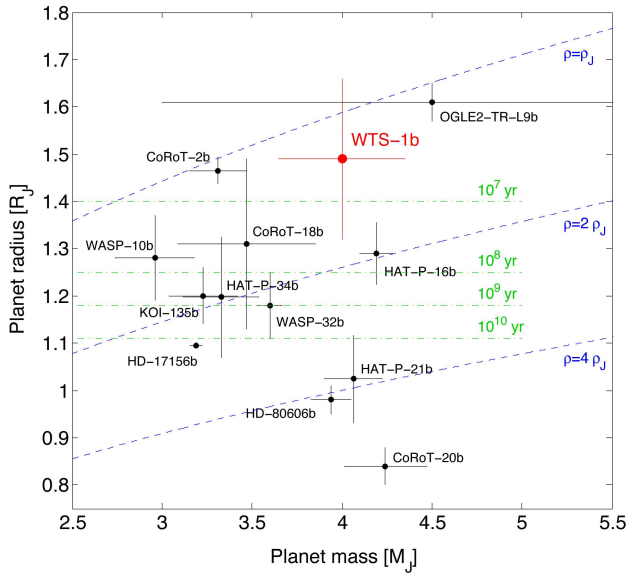}
\caption{Mass -- Radius diagram of the known planets with a mass in the range
         3-5 \JM (black dots). Labels with the related planet name are shown for an easier 
         identification. 
         Planets with only an upper limit on mass and/or radius are 
         not shown. The blue dashed lines represent the iso-density curves.
         The green dot-dashed lines indicate the planetary radii at different ages 
         accordingly to \citet{Fortney07} (see text for details).
         Masses, radii of the planets are taken from {\it www.exoplanet.eu} at the 
         time of the publication of this work, while the related uncertainties were found 
         in the refereed publication. WTS-1b is shown in red.}
\label{plot_Mp_Rp}
\end{figure}

In this paper we announce the discovery of a new transiting extrasolar planet, WTS-1b,
the first detected by the UKIRT/WFCAM Transit Survey.
The parameters of the planet are collected in {\mbox{Table}}~\ref{tab_WTS1b}.
WTS-1b is a $\sim$4 \JM\ planet orbiting in 3.35 days a late F-star with possibly slightly 
subsolar metallicity.
With a radius of \Grad\ \JR, it is located in the upper part of the mass -- radius 
diagram of the known extrasolar planets in the mass range 3-5 \JM\ (see
{\mbox{Figure}}~\ref{plot_Mp_Rp}).
The parameters of the other planets are taken from {\it www.exoplanet.eu} at the time 
of the publication of this work. Planets with only an upper limit on the mass and/or on 
the radius are not shown.

It is worth noting that only a cut-off of the transit depth, different for each survey, 
could act as a selection effect against the detection of planets in this upper portion of 
the diagram.Larger planetary radii imply a deeper transit feature 
in the light curves and thus, within this mass range, larger object are more easily 
detectable. The properties of WTS-1b, as well as those of the other two planets present in 
the upper part of the diagram, CoRoT-2b \citep{Alonso08} and OGLE2-TR-L9b \citep{Snellen09}, 
are not explained within standard formation and evolution models of isolated gas giant 
planets \citep{Guillot05}.

The radius anomaly is at the $\sim2\sigma$ level considering the stellar irradiation
that retards the contraction of the planets, the distance of the planet from the host 
star and the age of the planet \citep{Fortney07}. The models of Fortney and collaborators 
predict indeed a radius of 1.2 \JR\ for a 600\,Myr-old planet 
(see their Figure 5, 3\,\JM\ and 0.045\,AU model). 
This radius estimate is an upper limit as 600\,Myr is the lower limit on the 
age of the WTS-1 system due to the Li abundance (see Section~\ref{wts1}).
The radius trend shown in the figure would suggest an age for WTS-1b less than 10\,Myr.
The same significance on the radius anomaly is obtained considering empirical relationships
coming from the fit of the observed radii as a function of the physical properties of the 
star-planet system, such as mass, equilibrium temperature and tidal heating 
\citep[][eq. 10]{Enoch12}. In any case, a rapid migration of WTS-1b inward to the 
highly-irradiated domain after its formation seems required.

Surface day/night temperature 
gradients due to the strong incident irradiation, are likely to generate strong wind 
activity through the planet atmosphere. Recently, \citet{WuLithwick12} showed how the 
Ohmic heating proposal \citep{BatyginStevenson10,Perna10}
can effectively bring energy in the interior of the planet and slow down the cooling 
contraction of a HJ even on timescales of several Gyr:
a surface wind blowing across the 
planetary magnetic field acts as a battery that rises Ohmic dissipation in the deeper layers.
In \citet{Huang12}, the Ohmic dissipation in HJs is treated decoupling the interior of 
the planet and the wind zone. In this scenario, the radius evolution for an irradiated 
HJ planet (see their Figure 9, 3\,\JM) leads to a value consistent with our observation 
up to 3\,Gyr.

Accordingly to \citet{Fortney08}, the incident flux (the amount of energy from the host 
star irradiation, per unit of time 
and surface, that heats the surface of the planet) computed for WTS-1b 
(1.12$\pm$0.26 $\cdot$ 10$^{9}$\,erg~s$^{-1}$~cm$^{-2}$) assigns it to the so called pM class 
of HJs. This classification considers the day-side atmospheres of the highly-irradiated HJs 
that are somewhat analogous to the M- and L-type dwarfs. 
In particular, the predictions of equilibrium chemistry for pM planet atmospheres are similar
to M-dwarf stars, where absorption by TiO, VO, $H_2O$, and CO is prominent \citep{Lodders02}.
Planets in this class are warmer than required for condensation of titanium (Ti)- and
vanadium (V)-bearing compounds and will possess a temperature inversion (which could 
lead to a smaller inflation due to Ohmic heating accordingly to \citet{Heng12}) 
due to absorption of incident flux by TiO and VO molecules.
\citet{Fortney08} propose that these planets will have large day/night effective 
temperature contrasts and an anomalous
brightness in secondary eclipse at mid-infrared wavelengths. Unfortunately, the SNR in the 
$J$-band light curve, due to the faintness of the parent star WTS-1, is not high enough for 
such kind of detection (see Section~\ref{transit}).

To conclude, the discovery of WTS-1b demonstrates the capability of WTS to find 
planets, even if it operates in a back-up mode during dead time on a queue-schedule telescope  
and despite of the somewhat randomised observing strategy.
Moreover, WTS-1b is an inflated HJ orbiting a late F-star even if the project is designed 
to search for extrasolar planets hosted by M-dwarfs. 
\citealt{Birkby12b} will present the second WTS detection, WTS-2b, a Jupiter-like planet 
around a cool K-star.

\section*{Acknowledgments} 

We acknowledge support by RoPACS during this research, a Marie Curie Initial 
Training Network funded by the European Commissions Seventh Framework Programme.
The United Kingdom Infrared Telescope is operated by the Joint Astronomy Centre on 
behalf of the Science and Technology Facilities Council of the U.K.; some of the data 
reported here were obtained as part of the UKIRT Service Programme.
The Hobby-Eberly Telescope (HET) is a joint project of the University of Texas at Austin, 
the Pennsylvania State University, Stanford University, Ludwig-Maximilians-Universit\"at 
M\"unchen, and Georg-August-Universit\"at G\"ottingen. The HET is named in honor of its 
principal benefactors, William P. Hobby and Robert E. Eberly.
The 2.5m Isaac Newton Telescope and the William Herschel Telescope are operated on the 
island of La Palma by the Isaac Newton Group in the Spanish Observatorio del Roque de 
los Muchachos of the Instituto de Astrof\'isica de Canarias. 
We thank Calar Alto Observatory, the German-Spanish Astronomical Center, Calar Alto, 
jointly operated by the Max-Planck-Institut f\"{u}r Astronomie Heidelberg and the 
Instituto de Astrof\'{i}sica de Andaluc\'{i}a (CSIC), for allocation of director's 
discretionary time to this program.
We thank Kitt Peak National Observatory, National Optical Astronomy 
Observatory, which is operated by the Association of Universities for Research in 
Astronomy (AURA) under cooperative agreement with the National Science Foundation.
This publication makes use of VOSA, developed under the Spanish Virtual Observatory 
project supported from the Spanish MICINN through grant AyA2008-02156.
This research has been funded by Spanish grants AYA 2010--21161--C02--02, 
CDS2006--00070 and PRICIT--S2009/ESP--1496.
This work was partly funded by the Funda\c{c}\~ao para a Ci\^encia e a
Tecnologia (FCT)-Portugal through the project PEst-OE/EEI/UI0066/2011.
MC is grateful to JB for the cordial and fruitful collaboration.
MC thanks A. Driutti, B. Sartoris and P. Miselli for technical support and 
stimulating discussions.
NL was funded by the Ram\'on y Cajal fellowship number 08-303-01-02 and the
national program AYA2010-19136 funded by the Spanish ministry of science 
and innovation.
This work was co-funded under
the Marie Curie Actions of the European Commission (FP7-COFUND).
EM was partly supported by the CONSOLIDER-INGENIO
GTC project and the project AYA2011-30147-C03-03.
This research has made use of NASA's Astrophysics Data System.
We thank X. Chen and C. Sturm who helped us detecting the
mistake in Table 2.


\label{lastpage}

\end{document}